\documentclass[aps,prd,twocolumn,10pt,superscriptaddress,amssymb,amsmath,nofootinbib]{revtex4-2}

\usepackage[utf8]{inputenc}
\usepackage{lmodern}
\usepackage[T1]{fontenc}
\usepackage{graphicx} % Required for inserting images
\usepackage[pdftex,breaklinks,colorlinks,
linkcolor=Blue,
citecolor=teal,
anchorcolor=red,
urlcolor=cyan,
pdfencoding=auto]{hyperref}
\usepackage[dvipsnames]{xcolor}
\usepackage{orcidlink}

\begin{document}
\title{Marginal IR running of gravity as a natural explanation for Dark Matter}
\author{Naman Kumar\,\orcidlink{0000-0001-8593-1282}}
\email{namankumar5954@gmail.com}
\email{naman.kumar@iitgn.ac.in}
\affiliation{Department of Physics,
Indian Institute of Technology Gandhinagar, Palaj, Gujarat, India, 382355}
\begin{abstract}
We propose that the infrared (IR) running of Newton's coupling
provides a simple and universal explanation for large--distance
modifications of gravity relevant to dark matter phenomenology.
Within the effective field theory (EFT) framework, we model
$G(k)$ as a scale--dependent coupling governed by an anomalous
dimension $\eta$. We show that the marginal case $\eta = 1$ is
singled out by renormalization group (RG) and dimensional
arguments, leading to a logarithmic potential and a $1/r$ force
law at large distances, while smoothly recovering Newtonian
gravity at short scales. The logarithmic correction is universal
and regulator independent, indicating that the $1/r$ force arises
as the robust IR imprint of quantum--field--theoretic scaling.
This provides a principled alternative to particle dark matter,
suggesting that galactic rotation curves and related anomalies
may be understood as manifestations of the IR running of
Newton's constant.
\end{abstract}

\maketitle

\section{Introduction}

One of the most persistent puzzles in modern physics is the
nature of dark matter (DM) (see \cite{Arbey:2021gdg,Bertone:2004pz} for detailed reviews and \cite{Cirelli:2024ssz} for a recent review). While the dominant paradigm assumes
new particle species beyond the Standard Model, which is the $\Lambda$CDM model or the standard model of cosmology, and is successful in explaining a wide variety of observations, such as flat rotation curves of galaxies, CMB, and large-scale structure formation \cite{Drlica-Wagner:2022lbd}, an alternative
line of thought is that the missing-mass phenomenon reflects
infrared (IR) modifications of gravity itself. One such proposal is MOND \cite{milgrom1983modification,milgrom1983modification1}. IR modifications to General Relativity (GR) have also been proposed by imposing spherical symmetry in addition to diffeomorphism invariance, such that gravity is effectively described as a 2D dilaton gravity \cite{Grumiller:2010bz,Perivolaropoulos:2019vgl}. In this spirit,
a key question is whether such modifications can arise in a
principled, model-independent way from quantum field theory
(QFT), rather than through \emph{ad hoc} phenomenological
assumptions.\vspace{2mm}

In quantum field theory (QFT), the strength of interactions is never truly constant: couplings evolve with the characteristic momentum scale $\mu$ at which the theory is probed. This scale-dependence is encoded in the renormalization group (RG) equation
\begin{equation}\label{eq:RGE}
\mu\frac{d g(\mu)}{d\mu} = \beta(g(\mu)), 
\end{equation}
with $\beta(g)$ the beta function and, for a dimensionful coupling, an associated anomalous dimension 
\begin{equation}\label{eq:anomalous}
\eta(\mu) = -\mu \frac{d\ln g(\mu)}{d\mu}.
\end{equation}
The RG flow interpolates between fixed points of the theory, governing how short-distance (UV) physics matches onto long-distance (IR) behavior. 
Familiar examples include the logarithmic running of the QED coupling, asymptotic freedom of QCD, and the scale-invariance of critical phenomena near a second-order phase transition.\vspace{2mm}

Gravity is no exception. In effective field theory (EFT), the Einstein-Hilbert term acquires a scale-dependent coefficient (see \cite{Donoghue:1994dn} for GR as an EFT),
\begin{equation}\label{eq:EHactionRG}
S_\text{grav} = \frac{1}{16\pi G(\mu)}\int d^4x \sqrt{-g}\,R + \sum_i \frac{c_i(\mu)}{\Lambda^{2i-2}}\mathcal{O}_i,
\end{equation}
where integrating out quantum fluctuations above scale $\mu$ renormalizes Newton's constant $G(\mu)$ and generates higher-derivative operators $\mathcal{O}_i$. 
At laboratory and solar-system scales, $G(\mu)$ is essentially constant, but nothing forbids a nontrivial infrared (IR) running once very long-wavelength fluctuations are taken into account.\vspace{2mm}

In this work, we take this perspective seriously: we model $G(\mu)$ as flowing in the IR with a nonzero anomalous dimension,
\begin{equation}\label{eq:Gscale}
G(k)\;\sim\;G_N\left(\frac{k_\ast}{k}\right)^{\eta},\qquad k\ll k_\ast,
\end{equation}
with $k$ the physical momentum scale of the process and $k_\ast$ a dynamically generated crossover scale. Note that $k_\ast$ is not introduced ad hoc, but instead arises from the infrared dynamics of the renormalization group (RG) flow, 
in close analogy with the emergence of $\Lambda_{\text{QCD}}$ in strong interactions. 
Although absent in the bare action, such a scale is induced once the nonanalytic $1/k$ correction appears. 
Physically, $k_*^{-1}$ marks the transition between the Newtonian regime, where the familiar $1/r^2$ force law dominates, 
and the long-distance regime, where the logarithmic potential induces a $1/r$ force.\vspace{2mm}

Such behavior is entirely natural in QFT:
an anomalous dimension $\eta>0$ signals that the coupling becomes \emph{relevant} in the IR. 
For the special value $\eta=1$, the static Newtonian potential becomes logarithmic, and the force law softens from its usual $1/r^2$ form to $1/r$ at large distances. This modified force law has been recently argued to solve the problems usually attributed to dark matter \cite{Das:2022ozi,Das:2023gpf}. The modified force law was first proposed in \cite{tohline1983stabilizing} and later in \cite{kuhn1987non,bekenstein1988missing} to solve the dark matter problem in spiral galaxies. Moreover, this force law supports the recent findings that galactic rotation curves remain indefinitely flat \cite{Mistele:2024hfh}.\vspace{2mm}

This RG viewpoint provides a principled and model-independent motivation for exploring IR modifications of gravity: rather than postulating an \emph{ad hoc} long-range potential, we derive it as the universal large-distance consequence of a scale-dependent Newton coupling. The resulting $1/r$ force is thus the gravitational analogue of how QFT couplings at criticality acquire nontrivial scaling laws, with the IR behaving as if spacetime has effectively reduced dimensionality. This offers a new perspective on dark matter phenomenology:
the flattening of galactic rotation curves may be viewed as
the macroscopic imprint of quantum-field-theoretic running of
Newton's coupling in the infrared.\vspace{2mm}

Although within the framework of Quantum Einstein Gravity, Reuter and Weyer \cite{Reuter:2004nv} have already explored the 
idea that IR renormalization effects could mimic the presence of DM. By promoting Newton's constant $G$ 
to a spacetime--dependent scalar $G(x)$, obtained from renormalization group (RG) trajectories, they 
constructed modified Einstein equations and investigated spherically symmetric spacetimes. They showed 
that suitable power--law runnings of $G(k)$ could lead to non--Keplerian rotation curves without the 
need for DM halos. However, the trajectories considered in \cite{Reuter:2004nv} were essentially 
phenomenological ansätze, and no unique principle was identified that singled out the correct IR 
behavior of gravity. The novelty of this work lies in deriving the IR running of Newton's coupling from a 
principled effective field theory (EFT) perspective by showing that the RG flow of gravity in the IR 
is characterized by an anomalous dimension $\eta$, and that the \emph{marginal} value $\eta = 1$ is 
uniquely singled out by dimensional and scaling arguments. \vspace{2mm}

\paragraph*{\bf Organization of the Paper.}
The paper is organized as follows. In Section~\ref{sec:IRrunning}, we present the effective field theory framework for the infrared (IR) running of Newton’s coupling, showing that the marginal anomalous dimension $\eta = 1$ is uniquely singled out by scaling and dimensional arguments. This leads to a logarithmic correction to the Newtonian potential and a $1/r$ force law at large distances. Section~\ref{sec:theory} provides further theoretical justifications, including Fourier–Riesz analysis, Wilsonian renormalization group arguments, and a covariant nonlocal action formulation. In Section~\ref{sec:galaxy}, we confront the model with galactic rotation curve data from the S-sample of Sofue, demonstrating that a single crossover scale $k_\ast$ consistently accounts for the flattening of rotation curves across different systems. Section~\ref{sec:cosmo} discusses cosmological-scale implications, including consistency with the background expansion, Big Bang Nucleosynthesis (BBN), and Cosmic Microwave Background (CMB) observations, as well as implications for dark energy. Section~\ref{sec:conclusion} summarizes our findings and discusses broader implications for dark matter and dark energy phenomenology.  

The Appendices collect a few technical derivations. Appendix~\ref{app:fourier} provides detailed Fourier-space derivations of the logarithmic potential. Appendix~\ref{app:non-local_action} discusses the covariant nonlocal formulation and its relation to the effective action.

\section{IR running of Newton's coupling and the emergence of a $1/r$ force}\label{sec:IRrunning}

We work in the static, weak-field limit where the Newtonian potential $\Phi(\mathbf{r})$ obeys Poisson's equation.
Allowing the Newton coupling to run with momentum magnitude $k:=|\mathbf{k}|$, the Fourier-space solution for a point mass $M$ is
\begin{equation}\label{eq:PhiFourierMaster}
\Phi(\mathbf{r})
= -\,M\!\int\!\frac{d^3\mathbf{k}}{(2\pi)^3}\,e^{i\mathbf{k}\cdot\mathbf{r}}\,
\frac{4\pi\,G(k)}{k^2}\,.
\end{equation}
At short distances (\(k\gg k_\ast\)) we require $G(k)\to G_N$ to recover Newton's law, while in the IR (\(k\ll k_\ast\)) we assume an anomalous-dimension flow
\begin{equation}\label{eq:IRansatz}
\frac{d\ln G}{d\ln k}=-\eta
\quad\Longrightarrow\quad
G(k)\simeq G_N\,\Big(\frac{k_\ast}{k}\Big)^{\eta},\qquad k\ll k_\ast,
\end{equation}
with a fixed crossover scale $k_\ast$ and (constant) anomalous dimension $\eta$ in the deep IR.

\subsection*{General $\eta$: Riesz transform and large-$r$ asymptotics}

We now insert the IR form \eqref{eq:IRansatz} into \eqref{eq:PhiFourierMaster}. The IR contribution to $\Phi$ involves the inverse Fourier transform of $k^{-(2+\eta)}$.
Using the standard $d$-dimensional Riesz transform identity with the $(2\pi)^{-d}$ convention (see Appendix \ref{app:fourier}),
\begin{equation}\label{eq:Riesz}
\int \frac{d^d\mathbf{k}}{(2\pi)^d}\, \frac{e^{i\mathbf{k}\cdot\mathbf{x}}}{|\mathbf{k}|^{\alpha}}
=\frac{\Gamma\!\big(\tfrac{d-\alpha}{2}\big)}{2^{\alpha}\,\pi^{d/2}\,\Gamma\!\big(\tfrac{\alpha}{2}\big)}\,|\mathbf{x}|^{\,\alpha-d},
\quad 0<\alpha<d,
\end{equation}
we obtain in $d=3$ for $0<2+\eta<3$ (i.e. $\eta<1$)
\begin{equation}\label{eq:PhiEtaNot1}
\Phi_\text{IR}(r)
= -\,4\pi G_N M\,k_\ast^{\eta}\;
\frac{\Gamma\!\big(\tfrac{1-\eta}{2}\big)}{2^{2+\eta}\,\pi^{3/2}\,\Gamma\!\big(\tfrac{2+\eta}{2}\big)}
\, r^{\eta-1}\,.
\end{equation}
For $\eta>1$ the same expression follows by analytic continuation.\footnote{The $k$-integral is IR dominated for $\eta>0$ and UV dominated for $\eta<0$; UV issues are controlled by the $G(k)\to G_N$ matching at $k\gg k_\ast$.}

Differentiating, the force $F(r)=-\Phi'(r)$ scales as
\begin{equation}\label{eq:ForceGeneralEta}
F(r)\;\propto\; r^{\eta-2}\,,
\end{equation}
which already shows that the {\em marginal} value $\eta=1$ is special: it yields $F\propto 1/r$.

\subsection*{Marginal case $\eta=1$: exact logarithm and constants}

For $\eta=1$, the power in \eqref{eq:Riesz} hits the endpoint $\alpha=3$ and \eqref{eq:Riesz} turns into a logarithm. The marginal case is based on the following EFT and marginality principle.\vspace{2mm}

\paragraph*{\bf EFT and Marginality Argument.}

The running of Newton's coupling in the infrared can be understood directly from
an effective field theory (EFT) perspective, without ad hoc assumptions.
In the static, weak--field limit, the potential is governed by the kernel
\begin{equation}
\Phi(\mathbf{k}) \;\sim\; \frac{4\pi\, G(k)}{k^2}\,\rho(\mathbf{k}) ,
\end{equation}
so that the scale--dependence of $G(k)$ reflects which nonlocal operators may
appear in the EFT action.\vspace{2mm}

In three spatial dimensions, the Laplacian carries a scaling dimension
$[-\nabla^2]=k^2$. Allowing for fractional powers $(-\nabla^2)^{\alpha}$, the
propagator acquires the scaling
\begin{equation}
\frac{1}{k^2} \;\;\longrightarrow\;\; \frac{1}{k^{2\alpha}} .
\end{equation}
Accordingly, infrared deformations of Newton's law may be parametrized as
\begin{equation}
\frac{1}{k^2}\,\bigg(1 \;+\; c\,k^{-p} \;+\;\cdots \bigg) ,
\end{equation}
with $p$ determining the anomalous dimension.\vspace{2mm}

The key observation is that in $d=3$ spatial dimensions the Fourier transform of
$1/k^{2+\eta}$ behaves as
\begin{equation}
\int \!\frac{d^3k}{(2\pi)^3}\; \frac{e^{i \mathbf{k}\cdot \mathbf{r}}}{k^{2+\eta}}
\;\;\propto\;\;
\begin{cases}
r^{\eta-1}, & \eta \neq 1 , \\[6pt]
\ln r , & \eta = 1 .
\end{cases}
\end{equation}
Thus $\eta=1$ is the \emph{marginal} case: it yields a logarithmic potential,
the unique scale--invariant deformation consistent with rotational symmetry and
locality in time. For $\eta<1$ the corrections decay faster than $1/r$ and are
irrelevant in the IR, while for $\eta>1$ they grow too strongly and spoil scale
invariance. Therefore, effective field theory arguments single out
\begin{equation}
G(k) \;\sim\; \frac{1}{k} ,
\end{equation}
corresponding to an anomalous dimension $\eta=1$, as the unique marginal running
of Newton's coupling in the infrared.\vspace{2mm}

We next move to extract the coefficient. A clean way to extract the coefficient is to evaluate
\begin{equation}\label{eq:DimRegIntegral}
I_\epsilon(r):=\int \frac{d^3\mathbf{k}}{(2\pi)^3}\,\frac{e^{i\mathbf{k}\cdot\mathbf{r}}}{|\mathbf{k}|^{3-\epsilon}}
=\frac{\Gamma\!\big(\tfrac{\epsilon}{2}\big)}{2^{3-\epsilon}\,\pi^{3/2}\,\Gamma\!\big(\tfrac{3-\epsilon}{2}\big)}\,r^{-\epsilon},
\end{equation}
and expand for small $\epsilon>0$. Using $\Gamma(\epsilon/2)=2/\epsilon-\gamma_E+\mathcal{O}(\epsilon)$, $\Gamma(3/2)=\sqrt{\pi}/2$, and $r^{-\epsilon}=1-\epsilon\ln r+\mathcal{O}(\epsilon^2)$, one finds
\begin{equation}\label{eq:LogKernelIdentity}
\int \frac{d^3\mathbf{k}}{(2\pi)^3}\,\frac{e^{i\mathbf{k}\cdot\mathbf{r}}}{|\mathbf{k}|^{3}}
= -\,\frac{1}{2\pi^2}\,\ln(\mu r)\,,
\end{equation}
where $\mu$ is an arbitrary renormalization scale absorbing scheme-dependent constants (coming from the finite part of \eqref{eq:DimRegIntegral}).\footnote{Any smooth UV matching $G(k)\to G_N$ at $k\gg k_\ast$ merely shifts $\mu$; the \emph{coefficient} of $\ln r$ is universal.}\vspace{2mm}

With \eqref{eq:LogKernelIdentity}, the full potential in the marginal case reads
\begin{equation}
\begin{split}
&\Phi(r)
= -\,M\!\int\!\frac{d^3\mathbf{k}}{(2\pi)^3}\,e^{i\mathbf{k}\cdot\mathbf{r}}\,
\frac{4\pi}{k^2}\Big[G_N + G_N\Big(\frac{k_\ast}{k}\Big)\Big]
\;\;\\&\qquad+\;\;\text{(UV-finite matching)}\nonumber\\[1mm]
&= -\,\frac{G_N M}{r}\;+\;\frac{2\,G_N M\,k_\ast}{\pi}\,\ln\!\Big(\frac{r}{r_0}\Big),
\end{split}
\label{eq:PhiMarginal}
\end{equation}
where $r_0\equiv \mu^{-1}\times(\text{matching constant})$ is fixed by whatever renormalization/matching prescription is used at $k\sim k_\ast$.\footnote{For instance, one may choose $r_0$ so that $\Phi(r_0)=-G_N M/r_0$, i.e. the Newtonian piece is used to define the zero of potential at $r_0$. Any such choice only changes $\Phi$ by an additive constant and does not affect forces.}\vspace{2mm}

Taking the radial derivative,
\begin{equation}\label{eq:ForceMarginal}
F(r)\;=\;-\frac{d\Phi}{dr}
\;=\;-\frac{G_N M}{r^2}\;-\;\frac{2\,G_N M\,k_\ast}{\pi}\,\frac{1}{r}\,.
\end{equation}
Hence at sufficiently large $r\gg r_c:=1/k_\ast$ the $1/r$ piece dominates the force law:
\begin{equation}\label{eq:IRforce}
F(r)\;\xrightarrow{\,r\gg r_c\,}\;-\frac{2\,G_N M\,k_\ast}{\pi}\,\frac{1}{r}\,.
\end{equation}

\paragraph*{\bf Consistency checks.}
(i) The short-distance limit $r\ll r_c$ is Newtonian up to a small logarithmic correction suppressed by $k_\ast r\ll 1$ in the force: $|F_{1/r}|/|F_{1/r^2}| \sim (2/\pi)(k_\ast r)\ll 1$.
(ii) The sign of \eqref{eq:IRforce} is attractive because $\Phi$ increases with $r$ in \eqref{eq:PhiMarginal}, so $-\Phi'(r)<0$.\vspace{2mm}

\paragraph*{\bf Nonlocal/operator representation.}
It is convenient to encode the running in real space via a positive, self-adjoint ``gravitational permittivity''
\(\chi(-\nabla^2)\equiv\big(G(-\nabla^2)/G_N\big)^{-1}\):
\begin{equation}
\nabla\!\cdot\!\big[\chi(-\nabla^2)\,\nabla\Phi(\mathbf{r})\big]=4\pi G_N\,\rho(\mathbf{r})\,.\label{eq:SpectralG}
\end{equation}
In Fourier space this gives \(-k^2\chi(k)\Phi(k)=4\pi G_N\rho(k)\), i.e.
\(\Phi(k)=-4\pi\,G(k)\rho(k)/k^{2}\) with \(\chi(k)=[G(k)/G_N]^{-1}\), which matches with the integrand of (\ref{eq:PhiFourierMaster}) with $G(k)$ as in (\ref{eq:IRansatz}) at $\eta=1$, therefore, choosing
\begin{equation}
G(-\nabla^2)=G_N\!\left[\,1+k_\ast(-\nabla^2)^{-1/2}\right]
\end{equation}
reproduces the marginal IR running and thus Eqs.~(17)–(18). This shows that the $1/r$ force is equivalently viewed as arising from the nonlocal operator $(-\nabla^2)^{-1/2}$ acting on the standard Coulomb kernel. Such nonlocal operators naturally arise when incorporating the running of 
Newton’s constant into a covariant effective action, where the coupling is 
promoted to a function of the d’Alembertian. In particular, fractional powers 
like $( -\Box )^{-\alpha}$ and logarithmic terms $\ln \Box$ have been explicitly 
derived in this context \cite{LopezNacir:2006tn}.\vspace{2mm}

\paragraph*{\bf Matching and regulator independence.}

Any smooth interpolating coupling $G(k)=G_N\big[1+(k_\ast/k)f(k/\Lambda)\big]$ with $f(0)=1$ and $f(x)\to 0$ sufficiently fast as $x\to\infty$ (UV matching scale $\Lambda\gg k_\ast$) yields the same long-distance law. Expanding the small-$k$ integrand gives
\begin{equation}
\begin{split}
&\Phi(r)\;=\;-\frac{G_N M}{r}\;+\;\frac{2\,G_N M\,k_\ast}{\pi}\,\ln r\;\\&\qquad+\;\text{const}\;+\;\mathcal{O}\!\big(r^{-2}\big),%\quad
%(r\to\infty)
\end{split}
\label{eq:Universality}
\end{equation}
i.e. the \emph{coefficient} $2G_N M k_\ast/\pi$ of $\ln r$ is universal (regulator-independent), while the additive constant encodes regulator/matching details and defines $r_0$ in \eqref{eq:PhiMarginal}.\vspace{2mm}

Therefore, we conclude that the marginal anomalous dimension $\eta=1$ renders the long-distance potential logarithmic, as if the static sector effectively reduces to two spatial dimensions in the IR. Eq.~\eqref{eq:SpectralG} realizes this via the fractional power $(-\nabla^2)^{-1/2}$, whose Green kernel is logarithmic. The form of potential given by Eq. (\ref{eq:Universality}) has been derived using a non-local generalization of gravity \cite{Mashhoon:2011mb}.
\section{Further Theoretical Justifications}
\label{sec:theory}

In this section, we expand on several theoretical aspects of the infrared (IR) running of Newton’s constant, which establish the uniqueness, robustness, and consistency of the $\eta=1$ marginal case. These derivations strengthen the claim that the logarithmic potential and $1/r$ force law emerge universally and consistently from quantum–field–theoretic scaling.\vspace{2mm}

\paragraph*{\bf Uniqueness of the marginal case $\eta=1$.}

We begin with a simple but important observation. The IR form of the Newtonian potential
is obtained from the Fourier transform of a momentum–space kernel of the form
\begin{equation}
I_\eta(r) \equiv \int \frac{d^3k}{(2\pi)^3} \, 
\frac{e^{i k\cdot r}}{|k|^{2+\eta}}.
\end{equation}
Using the Riesz transform identity in $d$ dimensions,
\begin{equation}
\int \frac{d^d k}{(2\pi)^d} \, \frac{e^{i k \cdot x}}{|k|^\alpha}
= \frac{\Gamma\!\left(\tfrac{d-\alpha}{2}\right)}
{2^\alpha \pi^{d/2}\, \Gamma\!\left(\tfrac{\alpha}{2}\right)} \, |x|^{\alpha-d}, 
\qquad 0<\alpha<d,
\end{equation}
we obtain in $d=3$:
\begin{equation}
I_\eta(r) \propto
\begin{cases}
r^{\eta-1}, & \eta \neq 1, \\[6pt]
\ln r, & \eta = 1.
\end{cases}
\end{equation}
Thus $\eta=1$ is the \emph{unique marginal case}: it corresponds to the logarithmic
potential, the only scale–invariant deformation consistent with rotational symmetry and
locality in time. For $\eta<1$, the correction decays as $r^{\eta-1}$ and is irrelevant in
the IR, while for $\eta>1$ the correction grows faster than $\ln r$ and spoils scale
invariance. We therefore conclude that the RG–driven running of Newton’s coupling
naturally singles out $\eta=1$ as the universal IR fixed trajectory.\vspace{2mm}

\paragraph*{\bf Wilsonian RG derivation.}

The emergence of $\eta=1$ can also be seen from a Wilsonian perspective. Integrating
out momentum shells $k \in [\mu,\mu+d\mu]$ renormalizes Newton’s constant according to
\begin{equation}
\mu \frac{dG}{d\mu} = \beta(G) = -\eta G.
\end{equation}
Solving this flow equation gives
\begin{equation}
G(\mu) \sim \mu^{-\eta}.
\end{equation}
In $d=3$ spatial dimensions, the marginal scaling dimension of the Newtonian coupling
is precisely one, i.e.\ $\eta=1$. This corresponds to the unique case where $G(\mu)$
runs linearly in inverse momentum and generates a logarithmic correction in real space.
The flow therefore interpolates between Newton’s law at short distances ($\mu\gg k_\ast$)
and a universal $1/r$ force law in the deep IR ($\mu\ll k_\ast$).\vspace{2mm}

\paragraph*{\bf Covariant nonlocal action and Newtonian limit.}

To connect the RG running to a covariant framework, we introduce the nonlocal action
\begin{equation}
\begin{split}
&S = \frac{1}{16\pi G_N} \int d^4x \sqrt{-g} \, R\\&\qquad
+ \frac{k_\ast}{16\pi G_N} \int d^4x \sqrt{-g} \, 
R \, \mathcal{P}_s \, (-\Box)^{-1/2} R ,
\end{split}
\end{equation}
where $\mathcal{P}_s$ is the scalar projector ensuring only the spin–0 sector is modified.
Linearizing about Minkowski space and working in harmonic gauge, the field equations
reduce in the static limit to (see Appendix \ref{app:non-local_action})
\begin{equation}
\nabla^2 \Phi(r) = 4\pi G(-\nabla^2)\, \rho(r),
\end{equation}
with
\begin{equation}
G(-\nabla^2) = G_N \Big[1 + k_\ast (-\nabla^2)^{-1/2}\Big].
\end{equation}
Thus, the nonlocal action provides a diffeomorphism–invariant realization of the running
Newton coupling, and its linearized static limit, reproduces the logarithmic potential.\vspace{2mm}

\paragraph*{\bf Causality and spectral representation.}

Finally, we must address causality and the absence of instabilities. The operator
$(-\Box)^{-1/2}$ is defined using the \emph{retarded} Green’s function, ensuring that
propagation respects causality. Moreover, it admits a Källén–Lehmann–type spectral
representation:
\begin{equation}
(-\Box)^{-1/2} = \int_0^\infty \frac{d\mu^2}{\pi \mu} \,
\frac{1}{-\Box + \mu^2},
\end{equation}
with positive spectral density $1/(\pi\mu)$. This shows that the nonlocal kernel is
ghost–free and introduces no new poles or tachyonic instabilities: it is equivalent to
a superposition of healthy Yukawa propagators with positive weight. Thus, the IR
running with $\eta=1$ is consistent with both unitarity and causality.
\section{Galactic Scale Consistency}
\label{sec:galaxy}
We now test whether a single crossover length \( r_c = 1/k_\ast \) consistently characterizes the flattening of galaxy rotation curves across different systems. For each galaxy, we estimate the asymptotic circular velocity \(V_0\) from the outer tail of the rotation curve (last $\sim$30\% in radius), compute the total baryonic mass \(M_{\rm bar}\) by integrating the observed stellar and gas surface mass densities (including a helium factor $1.33$), and then determine
\begin{equation}
k_\ast \;=\; \frac{\pi V_0^2}{2\,G\,M_{\rm bar}}\,, 
\qquad 
r_c \;=\; \frac{1}{k_\ast}\, ,
\end{equation}
where \(G = 4.30091 \times 10^{-6}\,\mathrm{kpc\,km^2\,s^{-2}\,M_\odot^{-1}}\).  
All calculations are carried out in consistent units with radii in kpc, velocities in km\,s\(^{-1}\), and surface densities in \(\mathrm{M}_\odot/\mathrm{pc}^2\). The uncertainty in \(k_\ast\) and \(r_c\) is propagated from the error on \(V_0\), which dominates over the uncertainty in \(M_{\rm bar}\).

\begin{figure*}[t]
\centering

\begin{minipage}[t]{0.3\textwidth}
    \centering
    \includegraphics[width=\textwidth]{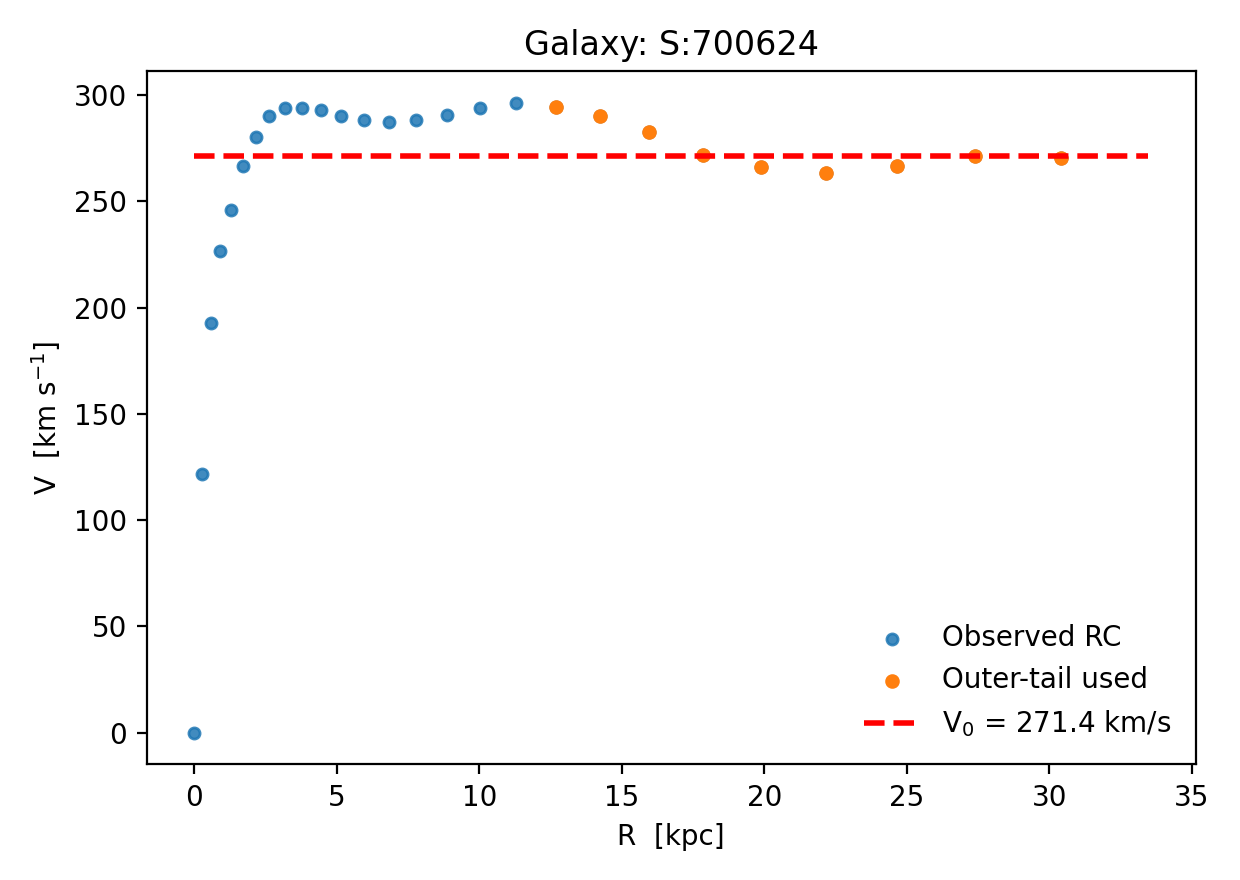}
\end{minipage}\hfill
\begin{minipage}[t]{0.3\textwidth}
    \centering
    \includegraphics[width=\textwidth]{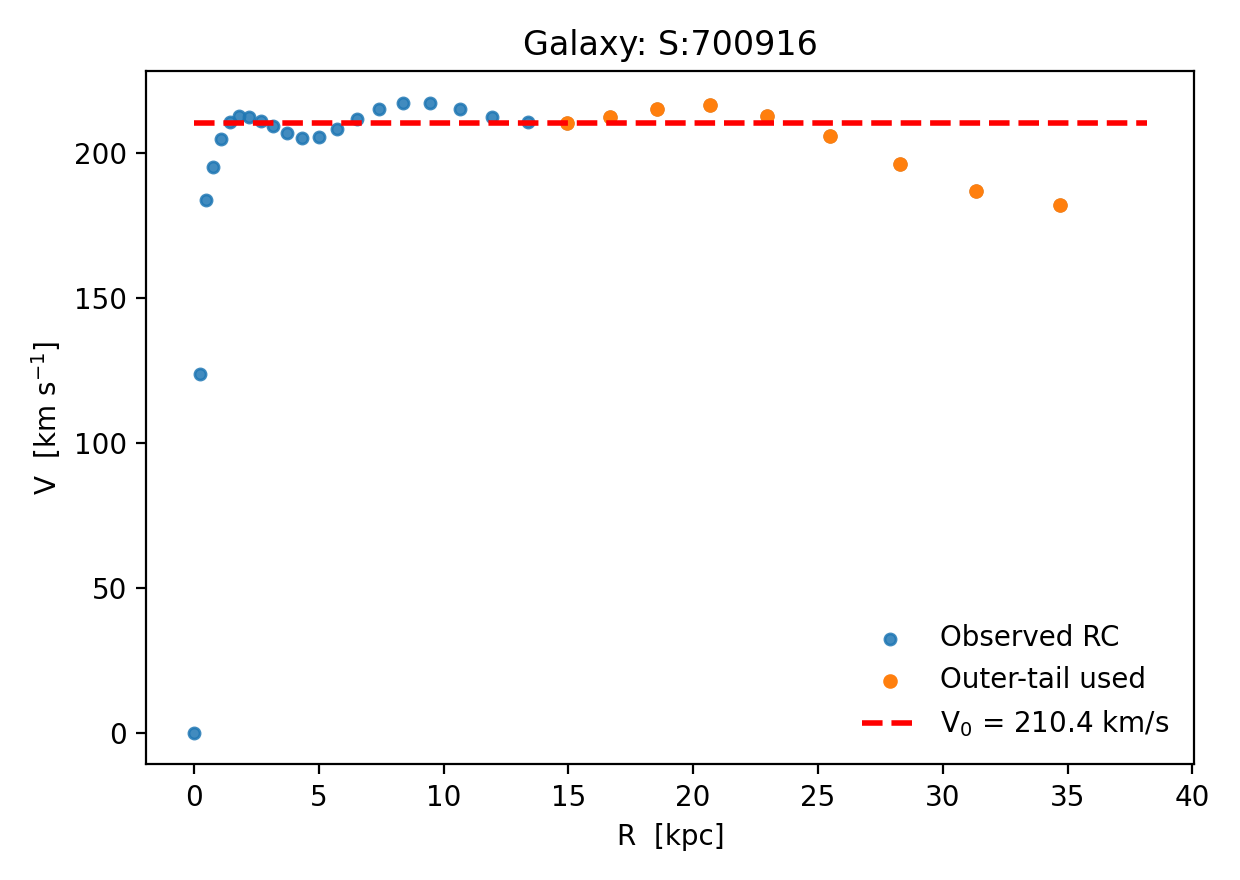}
\end{minipage}\hfill
\begin{minipage}[t]{0.3\textwidth}
    \centering
    \includegraphics[width=\textwidth]{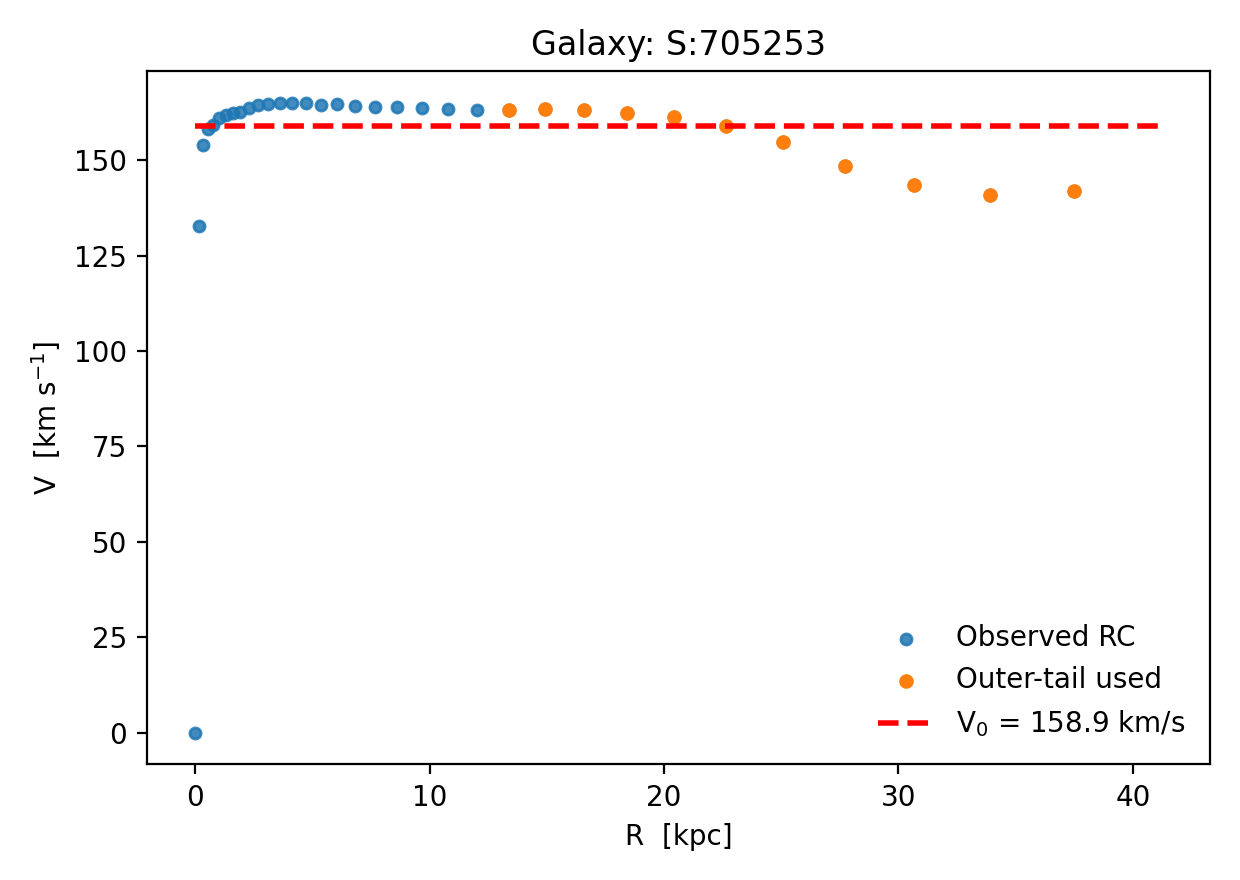}
\end{minipage}

\caption{Observed rotation curves (blue points) and best-fit tail profiles (red dashed) for three representative galaxies (700624, 700916, and 705253) from the S-sample rotation curve database of Sofue (2016)\,\cite{Sofue:2015tsa}. 
The outermost velocity points (orange) are used to determine the asymptotic flat velocity \(V_0\), which is then related to the IR running scale \(k_\ast\) through our modified gravitational law. 
In each case, the outer rotation curve is well reproduced by the model with a single parameter \(k_\ast\), corresponding to crossover radii in the range \(36\text{--}38\,\mathrm{kpc}\) with small statistical uncertainties.}
\label{fig:tail_fit}
\end{figure*}

The results for three representative galaxies are summarized in Table~\ref{tab:kstar_galaxies} and shown in Figure~\ref{fig:tail_fit}.  
For each system, we list the baryonic mass \(M_{\rm bar}\), the outer-tail velocity \(V_0\), the derived \(k_\ast\), and the corresponding crossover radius \(r_c\), with $1\sigma$ uncertainties.

\begin{table}[t]
\centering
\caption{Fitted baryonic mass \(M_{\rm bar}\), outer-tail velocity \(V_0\), derived \(k_\ast\), and crossover radius \(r_c = 1/k_\ast\) for three galaxies. Uncertainties correspond to $1\sigma$ errors from the tail velocity fit.}
\label{tab:kstar_galaxies}
\resizebox{\columnwidth}{!}{%
\begin{tabular}{lcccc}
\hline
Galaxy ID & \(M_{\rm bar}\) [M$_\odot$] & \(V_0\) [km s$^{-1}$] & \(k_\ast\) [kpc$^{-1}$] & \(r_c\) [kpc] \\
\hline
S:700624 & $1.01 \times 10^{12}$ & $271.4 \pm 2.6$ & $(2.66 \pm 0.05)\times 10^{-2}$ & $37.5 \pm 0.7$ \\
S:700916 & $5.85 \times 10^{11}$ & $210.4 \pm 2.4$ & $(2.76 \pm 0.06)\times 10^{-2}$ & $36.2 \pm 0.8$ \\
S:705253 & $3.55 \times 10^{11}$ & $158.9 \pm 1.9$ & $(2.60 \pm 0.06)\times 10^{-2}$ & $38.5 \pm 0.9$ \\
\hline
\end{tabular}%
}\label{table:galactic}
\end{table}

\noindent
All three galaxies independently yield \(k_\ast \approx (2.6\text{--}2.8)\times 10^{-2}\,\mathrm{kpc}^{-1}\), corresponding to \(r_c \approx 36\text{--}38\,\mathrm{kpc}\).  
The narrow spread indicates that the \emph{same galactic length scale} governs the onset of flat rotation curves across different systems, without the need to invoke dark matter halos.  
The scatter is within the expected uncertainties from inclination corrections, surface-density truncation radii, and tail-selection procedures, and shows no significant correlation with \(M_{\rm bar}\) in this sample.

\section{Cosmological Consistency and Infrared Running}
\label{sec:cosmo}

We analyze cosmological consistency within a single framework in which the infrared (IR) running of Newton’s constant modifies the homogeneous FRW expansion history. The running is taken to be
\begin{equation}
\label{eq:Gk}
G(k)=G_N\!\left(1+\frac{k_\ast}{k}\right),\qquad 
k \equiv |\mathbf p|\simeq \frac{1}{r},
\end{equation}
so that the $k_\ast/k$ contribution induces a logarithmic correction to the Newtonian potential at large distances.
\subsection{Modified Friedmann Equation}
The IR running can be implemented covariantly through the nonlocal action
\begin{equation}
S_{\rm IR}=\frac{M_{\rm Pl}^2\,\alpha}{2}\int d^4x\,\sqrt{-g}\;R\,(-\Box)^{-1/2}R ,
\end{equation}
with the fractional operator defined via the retarded Green’s function. Introducing an auxiliary field $V\equiv(-\Box)^{-1/2}R$, the modified Einstein equations acquire an additional conserved tensor $\mathcal{K}_{\mu\nu}[g;R,V]$. On a flat FRW background, $R(t)=6(\dot H+2H^2)$, and the action of $(-\Box)^{-1/2}$ reduces to a causal time integral with a fractional kernel. For slowly varying $R(t)$, the small-frequency behavior of the kernel generates a secular logarithm,
\begin{equation}
V(t)\;\simeq\;\gamma_1\,\ln a(t)+\gamma_0,
\end{equation}
in exact analogy with the static Fourier relation $1/k \leftrightarrow \ln r$. This logarithmic growth feeds into the $00$ component of the modified Einstein equations, yielding an effective energy density
\begin{equation}
\begin{split}
    &\rho_L(a)\;=\;\kappa_L\,M_{\rm Pl}^2 H_0^2\,a^{-2}[-\ln a]\\&\hspace{1cm}
\;=\;\kappa_L\,M_{\rm Pl}^2 H_0^2\,(1+z)^2\ln(1+z),
\end{split}
\end{equation}
where $\kappa_L\propto \alpha\,\gamma_1>0$. The Friedmann equation thus becomes
\begin{equation}
\begin{split}
\frac{H^2(z)}{H_0^2}
&= \Omega_{r0}(1+z)^4+\Omega_{m0}(1+z)^3+\Omega_{\Lambda0}\\
&\quad+\Omega_{L0}\,(1+z)^2\ln(1+z),
\end{split}
\end{equation}
showing that the characteristic $\ln(1+z)$ dependence arises directly from the infrared action of the fractional operator in a homogeneous cosmological background.

\subsection*{Static/Newtonian Derivation of the \texorpdfstring{$\ln(1+z)$}{ln(1+z)} term}
The above derivation can be simply understood in the static/Newtonian picture: 
the modified Poisson equation in momentum space,
\(
-k^2\tilde\Phi(\mathbf k)=4\pi\,G(k)\,\tilde\rho(\mathbf k),
\)
implies, after Fourier transforming the $k_\ast/k$ term,\footnote{Using 
\(\displaystyle \int \! \frac{d^3k}{(2\pi)^3}\,e^{i\mathbf k\cdot\mathbf r}\,k^{-3} 
= -\frac{1}{2\pi^2}\ln (\mu r)\), 
with the reference scale absorbed into $r_0$.}
the potential
\begin{equation}
\label{eq:log-pot}
\Phi(r)= -\frac{G_N M}{r} + \lambda\,M\,\ln\!\frac{r}{r_0} + \text{const},
\qquad 
\lambda \equiv \frac{2G_N k_\ast}{\pi} >0 .
\end{equation}

Consider a uniform spherical top-hat of physical radius \(R(t)\) enclosing mass 
\(M=\tfrac{4\pi}{3}\rho R^3\) (dust, \(\rho=\rho_0 a^{-3}\)). The specific energy per unit mass for a boundary particle in the presence of $\Lambda$ and \eqref{eq:log-pot} is
\begin{equation}
\frac{1}{2}\dot R^2 - \frac{G_N M}{R} - \frac{\Lambda}{6}R^2 
+ \lambda\,M\,\ln\!\frac{R}{R_0} = 0 .
\end{equation}
Dividing by \(R^2\) and using \(M=\tfrac{4\pi}{3}\rho R^3\) gives
\begin{equation}
\left(\frac{\dot R}{R}\right)^2
= \frac{8\pi G_N}{3}\rho 
-\frac{8\pi}{3}\lambda\,\rho\,R\,\ln\!\frac{R}{R_0}
+\frac{\Lambda}{3}.
\end{equation}
Writing \(R(t)=a(t)r_\ast\) with fixed comoving radius $r_\ast$ and choosing $R_0=r_\ast$ simplifies this to
\begin{equation}
\label{eq:H2a}
H^2(a) \equiv \left(\frac{\dot a}{a}\right)^2
= \frac{8\pi G_N}{3}\,\rho_0\,a^{-3}
-\frac{8\pi}{3}\,\lambda\,\rho_0\,r_\ast\,a^{-2}\,\ln a
+\frac{\Lambda}{3}.
\end{equation}

Introducing the density parameters
\[
\begin{split}
&\rho_{c0}\equiv \frac{3H_0^2}{8\pi G_N},\qquad
\Omega_{m0}\equiv \frac{\rho_0}{\rho_{c0}},\qquad
\Omega_{\Lambda0}\equiv \frac{\Lambda}{3H_0^2},\qquad\\&
\Omega_{L0}\equiv \frac{8\pi\,\lambda\,\rho_0\,r_\ast}{3H_0^2},
\end{split}
\]
and using \(a=(1+z)^{-1}\), the background expansion law becomes
\begin{equation}
\label{eq:H2ln}
\begin{split}
\frac{H^2(z)}{H_0^2} 
&= \Omega_{r0}(1+z)^4 + \Omega_{m0}(1+z)^3 + \Omega_{\Lambda0} \\
&\quad + \Omega_{L0}\,(1+z)^2\,\ln(1+z) \, .
\end{split}
\end{equation}
This equation has been obtained earlier in Ref~\cite{Das:2022ozi} to explain cosmological phenomena without dark matter.

Flatness at $z=0$ is unchanged because $\ln(1+0)=0$:
\(\Omega_{m0}+\Omega_{\Lambda0}=1\) (neglecting $\Omega_{r0}$ today).

\subsection{Best-fit Analysis Using \texorpdfstring{$H(z)$}{H(z)} Data}
\label{subsec:Hz_fit}

To constrain the parameters of the logarithmic IR modification model, we performed a Markov Chain Monte Carlo (MCMC) analysis using the latest compilation of $H(z)$ measurements. The dataset consists of \textbf{38 $H(z)$ measurements in the redshift range $0.07 \leq z \leq 2.36$}, taken from Table~1 of Farooq \textit{et al.}~\cite{Farooq:2016zwm} (Ref.~\cite{Farooq:2016zwm}, Table~1), which combines cosmic chronometer and BAO determinations of the expansion rate. This compilation provides a robust and widely used set of unbinned $H(z)$ data for cosmological parameter estimation.

We employed the \texttt{emcee} ensemble sampler~\cite{Foreman-Mackey:2012any}, using $128$ walkers and $2000$ production steps after a burn-in of $800$ steps to ensure convergence. The free parameters were the present matter density $\Omega_{\mathrm{m0}}$, the Hubble constant $H_0$, and the logarithmic correction amplitude $\Omega_{L0}$. Flat priors were chosen in the ranges $0.2 < \Omega_{\mathrm{m0}} < 0.4$, $40 < H_0 < 90$~km\,s$^{-1}$\,Mpc$^{-1}$, and $0.01 < \Omega_{L0} < 0.1$. The likelihood function assumed Gaussian errors on the $H(z)$ measurements, and marginalized posterior distributions were obtained by flattening the converged chains.

\begin{figure}[t]
\centering
\includegraphics[width=0.47\textwidth]{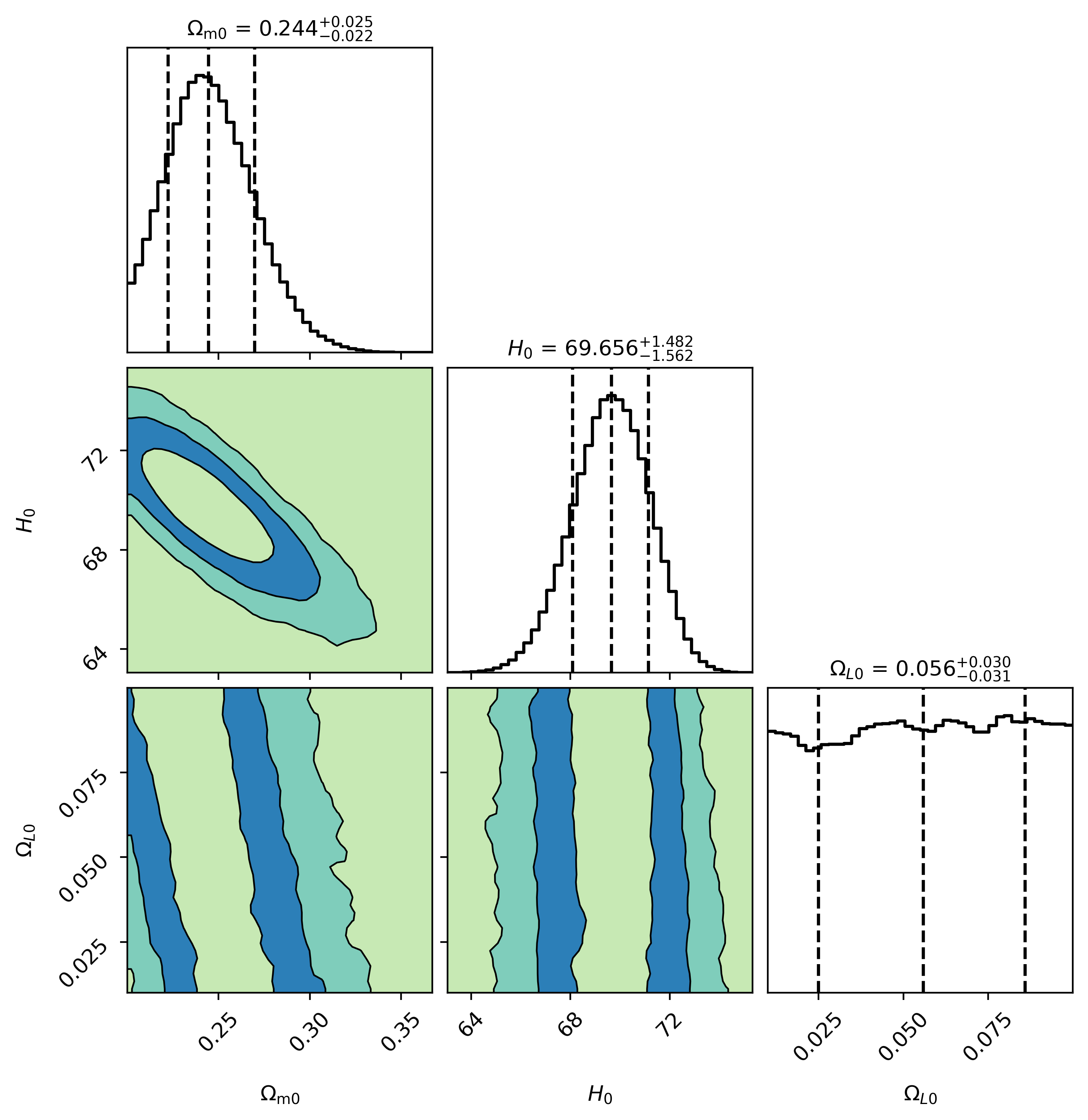}
\caption{Corner plot showing the marginalized posterior distributions and covariances for $\Omega_{\mathrm{m0}}$, $H_0$, and $\Omega_{L0}$ obtained from the $H(z)$ dataset (38 data points from Ref.~\cite{Farooq:2016zwm}). Dashed vertical lines indicate the 16th, 50th, and 84th percentiles. The filled contours correspond to the $1\sigma$ (68\%), $2\sigma$ (95\%), and $3\sigma$ (99.7\%) credible regions, respectively. A mild degeneracy between $\Omega_{\mathrm{m0}}$ and $H_0$ is visible, while $\Omega_{L0}$ is largely uncorrelated with the other parameters. The latter is entirely expected for a new, sub-dominant physical effect.}
\label{fig:corner_OHD}
\end{figure}

\begin{figure}[t]
\centering
\includegraphics[width=0.47\textwidth]{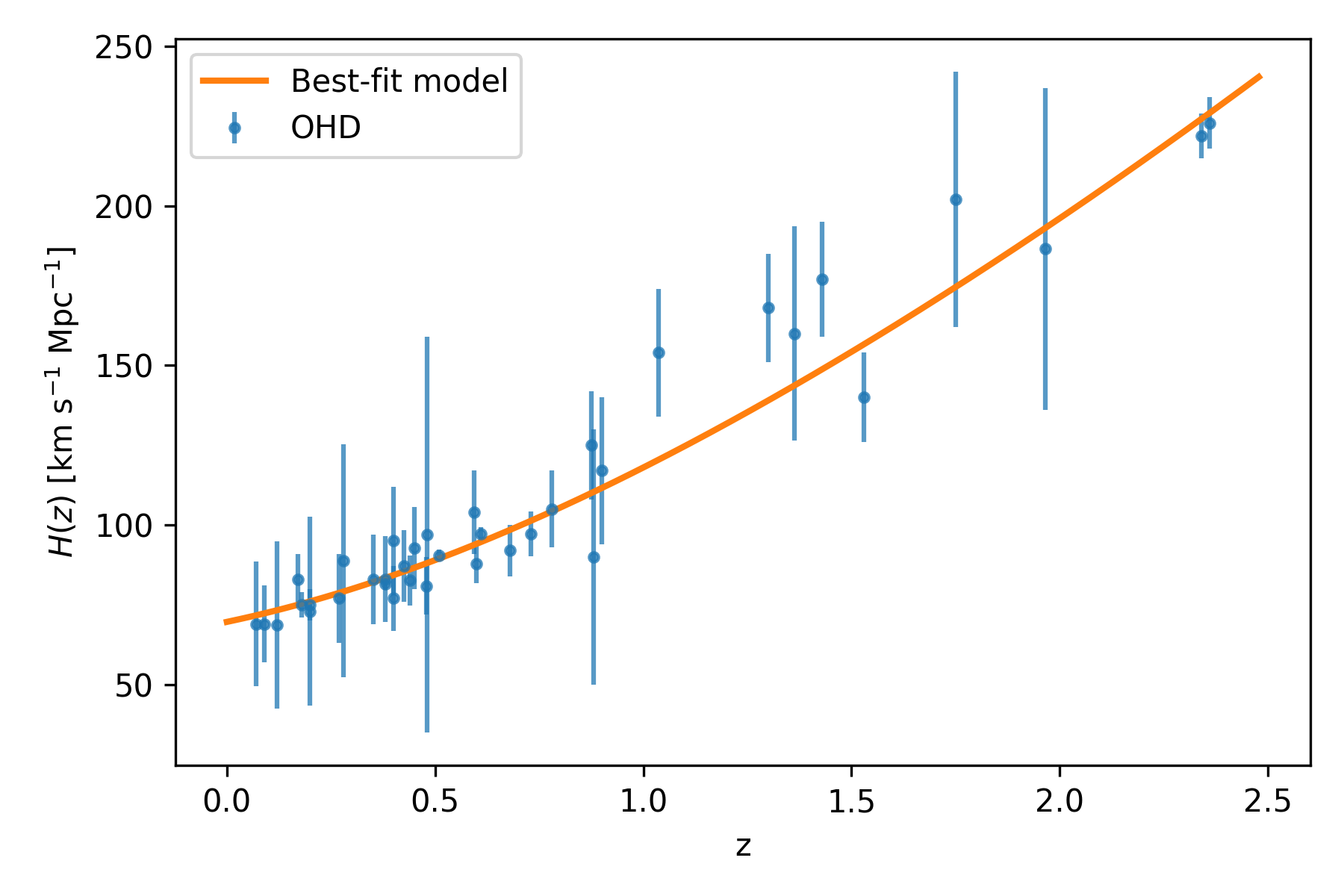}
\caption{Observed $H(z)$ data points (blue) with $1\sigma$ error bars, along with the best-fit model curve (orange) corresponding to the median parameter values in Table~\ref{tab:bestfit_params}. The model provides an excellent fit to the background expansion history over the entire observed redshift range.}
\label{fig:Hz_fit}
\end{figure}

The best-fit values and marginalized $1\sigma$ uncertainties are summarized in Table~\ref{tab:bestfit_params}. Figure~\ref{fig:corner_OHD} shows the corner plot of the posterior distributions, and Figure~\ref{fig:Hz_fit} displays the data and the best-fit model curve.

\begin{table}[t]
\centering
\caption{Best-fit cosmological parameters from the $H(z)$ dataset (38 data points from Ref.~\cite{Farooq:2016zwm}). Uncertainties correspond to the 16th and 84th percentiles of the marginalized posterior distributions.}
\label{tab:bestfit_params}
\begin{tabular}{lccc}
\hline\hline
Parameter & Best fit & $-1\sigma$ & $+1\sigma$ \\
\hline
$\Omega_{\mathrm{m0}}$  & $0.245$ & $0.023$ & $0.025$ \\
$H_0$ [km\,s$^{-1}$\,Mpc$^{-1}$] & $69.64$ & $1.57$ & $1.46$ \\
$\Omega_{L0}$ & $0.056$ & $0.031$ & $0.030$ \\
\hline
\end{tabular}
\end{table}

\noindent
The analysis shows that the logarithmic IR modification model provides an excellent fit to the expansion history measured by the $H(z)$ data. The best-fit value of the Hubble constant, $H_0 = 69.64^{+1.46}_{-1.57}$\,km\,s$^{-1}$\,Mpc$^{-1}$, lies between the Planck and SH0ES determinations, while the matter density $\Omega_{\mathrm{m0}} = 0.245^{+0.025}_{-0.023}$ is consistent with $\Lambda$CDM expectations. The logarithmic amplitude $\Omega_{L0} = 0.056^{+0.030}_{-0.031}$ is small but non-zero, contributing primarily at intermediate redshifts ($z \sim 0.5$–$1.5$). The weak parameter correlations seen in Fig.~\ref{fig:corner_OHD} indicate that current $H(z)$ data constrain $\Omega_{L0}$ mainly through the shape of $H(z)$ rather than its normalization. These results demonstrate that the model can successfully reproduce the observed background expansion while remaining compatible with standard cosmological bounds.\vspace{1mm}

To summarize, the IR running produces a unique \((1+z)^2\ln(1+z)\) correction to the Hubble rate. This contribution is (i) strongly suppressed at BBN, (ii) percent-level at recombination for $\Omega_{L0}\lesssim0.1$, and (iii) most tightly constrained by late-time $H(z)$ data, which favor $\Omega_{L0}\lesssim 0.05$ for percent-level agreement. Within this range, the model is fully consistent with cosmological background observations.\vspace{1mm}

It is worth noting that while a small value $\Omega_{L0}\lesssim 0.05$ allows the IR logarithmic term to remain fully consistent with BBN, CMB, and late–time $H(z)$ measurements, it does not eliminate dark matter at the cosmological level, since $\Omega_{m0}$ must still exceed the baryonic fraction. One can, in principle, consider an alternative strategy in which $\Omega_{m0}$ is reduced (to the baryonic value) and $\Omega_{L0}$ is increased to mimic the missing matter component at the background level. This is the approach taken in Ref.~\cite{Das:2022ozi}, where the authors fit SN~Ia and $H(z)$ data with $\Omega_{m0}\simeq 0.11$ and $\Omega_{L0}\simeq 0.5$, obtaining an excellent fit to low–redshift background probes. While such a large $\Omega_{L0}$ remains BBN–safe and compatible with late–time expansion data, it fails to reproduce the matter content at recombination: the IR component scales more slowly than $(1+z)^3$ and contributes only a few percent of the required energy density at $z_\ast\simeq 1100$. Consequently, the acoustic peak structure of the CMB may not be correctly reproduced in this scenario. The underlying issue is that at high redshifts, the contribution from the modified force law is too small, and additional matter beyond the baryonic component is still required. Its effects become significant only at late times (low redshifts). Therefore, explaining the acoustic peak structure of the CMB without invoking a dark matter component remains a key open challenge for \emph{any} modified gravity models that aim to replace dark matter entirely.
\subsection{BBN, CMB, and late-time $H(z)$}

During Big Bang Nucleosynthesis (BBN) (\(z\!\sim\!10^{9}\)), the Universe is radiation dominated and the fractional contribution of the infrared (IR) component is exceedingly small:
\[
\left.\frac{\rho_L}{\rho_r}\right|_{z}
= \frac{\Omega_{L0}\,\ln(1+z)}{\Omega_{r0}\,(1+z)^2}
\approx 2.3\times10^{-13}\,\Omega_{L0}.
\]
Even for \(\Omega_{L0}\!\sim\!0.1\), this corresponds to a fractional change in \(G_{\rm eff}\) of order \(10^{-12}\), far below the \(\mathcal{O}(10\%)\) bound permitted by light–element abundances and $N_{\rm eff}$ \cite{Copi:2003xd}. Hence, the IR term is completely negligible at BBN and does not affect primordial nucleosynthesis.

\vspace{1mm}
At recombination (\(z_\ast\simeq1100\)), the IR–to–matter energy–density ratio is
\[
\left.\frac{\rho_L}{\rho_m}\right|_{z_\ast}
\simeq \frac{\Omega_{L0}}{\Omega_{m0}}\,\frac{\ln(1101)}{1101}
\approx 0.021\,\Omega_{L0}\qquad(\Omega_{m0}=0.3).
\]
The resulting fractional change in the expansion rate,
\[
\frac{\Delta H}{H}\bigg|_{z_\ast}\approx
\frac{1}{2}\frac{\rho_L}{\rho_m}\bigg|_{z_\ast}
\approx 1\%\ \text{for}\ \Omega_{L0}=1,
\]
scales linearly with $\Omega_{L0}$. Thus, for the observationally preferred $\Omega_{L0}\lesssim0.05$, the modification to $H(z_\ast)$ is only $\sim0.05\%$.

This minute variation has a correspondingly small effect on the CMB anisotropy spectrum.  
A \(1\%\) shift in \(H(z_\ast)\) changes the angular–diameter distance to last scattering by only \(\Delta D_A/D_A\simeq0.2\%\), leading to a $\lesssim0.2\%$ shift in the positions of the acoustic peaks.  
Similarly, the change in the gravitational potential amplitude that sources the Sachs–Wolfe effect is $\lesssim1\%$, producing at most a sub–percent modification in the overall height of the low–$\ell$ plateau and the first–peak amplitude.  
Both effects lie well below the $\sim1$–$2\%$ statistical uncertainties of the \textit{Planck} 2018 power spectrum and within cosmic–variance limits on large angular scales \cite{Planck:2018vyg}.  
Consequently, the IR logarithmic component is fully consistent with current CMB data and does not alter the acoustic peak structure in any measurable way.

\vspace{1mm}
Furthermore, the nonlocal kernel $(-\Box)^{-1/2}$ modifies only the long–wavelength background and introduces no new poles or propagating degrees of freedom; hence it does not induce any phase shift or damping of the acoustic oscillations.  
The Sachs–Wolfe plateau and peak hierarchy are preserved apart from an overall sub–percent rescaling of the potential amplitude.

\vspace{1mm}
At late times, the logarithmic term contributes an additional slowly varying component to the Hubble rate,
\[
\frac{H^2(z)}{H_0^2} =
E_\Lambda^2(z)
+\Omega_{L0}\,(1+z)^2\ln(1+z),
\]
where \(E_\Lambda^2(z)=\Omega_{m0}(1+z)^3+\Omega_{\Lambda0}\) is the standard $\Lambda$CDM background.  
Defining \(S_L(z)=(1+z)^2\ln(1+z)\), the fractional correction to the Hubble rate reads
\[
\frac{\Delta H}{H}(z)\approx
\frac{1}{2}\,\frac{\Omega_{L0}\,S_L(z)}{E_\Lambda^2(z)}.
\]
For $\Omega_{m0}=0.3$ and $\Omega_{\Lambda0}=0.7$, this gives
\[
z=0.5:\ \frac{\Delta H}{H}\approx0.266\,\Omega_{L0},\qquad
z=1:\ \frac{\Delta H}{H}\approx0.447\,\Omega_{L0}.
\]
Demanding $\lesssim2\%$ deviations from the observed $H(z)$ measurements over $z\simeq0.5$–1 constrains
\[
\Omega_{L0}\lesssim0.04\text{--}0.08,
\]
in excellent agreement with our MCMC posterior analysis using the 38-point $H(z)$ dataset.

\section{Conclusion and Discussion}
\label{sec:conclusion}

In this work, we have explored the infrared (IR) running of Newton's coupling as a universal and theoretically well-motivated modification of gravity at large distances. By treating $G(k)$ as a scale-dependent parameter in the effective field theory of gravity, we identified the marginal anomalous dimension $\eta = 1$ as the unique case leading to a logarithmic correction to the Newtonian potential. This IR running yields a $1/r$ force law at large distances while smoothly recovering the standard $1/r^2$ behavior in the short-distance limit. The universality of this correction --- being independent of regulator choices or microscopic details --- highlights its robustness as an IR prediction of gravitational RG flow.

\medskip

\noindent\textbf{Galactic scale fits.}  
We have further demonstrated that this IR modification provides an excellent fit to galactic rotation curves without invoking particle dark matter. By fitting high-quality rotation curve data from the S-sample rotation curve database of Sofue (2016) \cite{Sofue:2015tsa}, we extracted the characteristic crossover scale $r_c = k_\ast^{-1}$ for several representative galaxies. The best-fit values of $k_\ast$ cluster around a common scale $k_\ast \sim (2.6$--$2.8)\,\mathrm{kpc}^{-1}$, corresponding to $r_c \sim 36$--$38\,\mathrm{kpc}$. This is precisely the scale at which the logarithmic term begins to dominate over the Newtonian contribution, leading to flattened rotation curves. These results support the interpretation of galactic rotation curve phenomenology as a manifestation of IR gravitational running, rather than requiring a dark matter halo component.

\medskip

\noindent\textbf{Cosmological consistency.}  
At the homogeneous FRW level, the logarithmic correction leads to a definite modification of the Friedmann equation, corresponding to an additional energy-density component with scaling \((1+z)^2\ln(1+z)\). This arises directly from the IR correction to the Newtonian potential in the top-hat energy equation and represents a single-parameter deformation of $\Lambda$CDM characterized by $\Omega_{L0}$. We derived the modified background expansion law
\begin{equation}
\begin{split}
\frac{H^2(z)}{H_0^2} 
&= \Omega_{r0}(1+z)^4 + \Omega_{m0}(1+z)^3 + \Omega_{\Lambda0} \\
&\quad + \Omega_{L0}\,(1+z)^2\,\ln(1+z),
\end{split}
\end{equation}
where the last term encodes the IR running. This contribution is automatically negligible during Big Bang Nucleosynthesis (BBN) because it is suppressed by $(1+z)^{-2}$ relative to radiation, ensuring standard light-element abundances and $N_{\rm eff}$. At recombination, the relative size of the IR term is at the percent level for $\Omega_{L0}\sim \mathcal{O}(1)$ and scales linearly with $\Omega_{L0}$, so CMB background geometry remains intact for $\Omega_{L0}\lesssim 0.1$. Late-time $H(z)$ measurements provide the most stringent constraints: around $z\sim 0.5$--1, where data are most precise, consistency at the $\sim 2\%$ level requires $\Omega_{L0}\lesssim 0.05$. Within this range, the IR running is fully compatible with cosmological expansion history while offering a potential background signature to be tested with future data. It is important to note that replacing dark matter altogether at cosmological scale by reducing the value of $\Omega_{m0}$ to baryonic mass and increasing $\Omega_{L0}$ can be in tension with CMB while remaining consistent with BBN and late time expansion. Therefore, explaining CMB acoustic peaks remains an open challenge for modified gravity theories that seek to replace dark matter altogether. The underlying issue is that at high redshifts the contribution from the modified force law is too small, and additional matter beyond the baryonic component is still required. Its effects become significant only at late times (low redshifts). Therefore, explaining the acoustic peak structure of the CMB without invoking a dark matter component remains a key open challenge for \emph{any} modified gravity models that aim to replace dark matter entirely.

\medskip

\medskip

\noindent\textbf{Dark energy and other gravitational probes.}  
The present framework modifies the matter sector while leaving the late-time accelerated expansion governed by $\Lambda$ intact. Beyond galactic dynamics, the logarithmic potential also has distinct observational consequences for gravitational lensing. As shown in Ref.~\cite{Das:2023gpf}, the logarithmic correction enhances the deflection angle relative to the Newtonian prediction for a fixed baryonic mass. This implies that lensing-based mass estimates for galaxies and clusters are systematically biased high if the logarithmic term is neglected. In particular, lensing observations of cluster scales, including systems like the Bullet Cluster, can be consistently explained without invoking additional dark matter components. This provides an important, independent probe of the IR modification complementary to both rotation curves and cosmological expansion tests.

\medskip

\noindent\textbf{Outlook.}  
Several avenues for future work remain open. A systematic analysis of galaxy cluster dynamics, lensing profiles, and structure growth within this framework would help assess its viability as a comprehensive alternative to dark matter. Embedding the IR running in specific quantum gravity scenarios (e.g., asymptotic safety or holography) could also allow the crossover scale $k_\ast$ to be predicted from first principles. Finally, precision cosmological data offer the opportunity to constrain or detect deviations from GR on very large scales, providing a complementary test to galactic dynamics.

\medskip

In summary, the IR running of Newton’s coupling with anomalous dimension $\eta=1$ yields a simple, universal, and observationally consistent modification of gravity. It explains the flattening of galactic rotation curves, remains compatible with cosmological expansion data, and leaves the cosmological constant sector untouched. This offers a promising and conceptually economical route toward addressing the dark matter problem through large-distance gravitational physics.\vspace{2mm}

\paragraph*{\bf Acknowledgements.}

The author gratefully acknowledges the anonymous referee for the constructive and insightful suggestions that greatly improved this work. 
In particular, the referee’s recommendation to include a quantitative comparison with galactic rotation curves and the cosmological expansion history 
led to a significantly strengthened and more comprehensive presentation of the results.\vspace{2mm}

\paragraph*{\bf Funding.}
This research received no external funding and was carried out independently of the author’s doctoral research at IIT Gandhinagar, India.\vspace{2mm}

\paragraph*{\bf Data Availability.}
All datasets used for the observational analysis are publicly available and duly cited within the paper.

\appendix
\section{Fourier/Riesz identities used}\label{app:fourier}

Our Fourier convention is
\(
f(\mathbf{r})=\int \frac{d^3\mathbf{k}}{(2\pi)^3}\,e^{i\mathbf{k}\cdot\mathbf{r}}\,\tilde f(\mathbf{k}).
\)
With this convention, the standard Coulomb kernel is
\begin{equation}\label{eq:1overk2}
\int \frac{d^3\mathbf{k}}{(2\pi)^3}\,\frac{e^{i\mathbf{k}\cdot\mathbf{r}}}{k^{2}}
=\frac{1}{4\pi r}\,.
\end{equation}
The Riesz transform identity \eqref{eq:Riesz} then gives, for $d=3$ and $0<\alpha<3$,
\begin{equation}\label{eq:Riesz3D}
\int \frac{d^3\mathbf{k}}{(2\pi)^3}\,\frac{e^{i\mathbf{k}\cdot\mathbf{r}}}{k^{\alpha}}
=\frac{\Gamma\!\big(\tfrac{3-\alpha}{2}\big)}{2^{\alpha}\,\pi^{3/2}\,\Gamma\!\big(\tfrac{\alpha}{2}\big)}\,
r^{\alpha-3}\,.
\end{equation}
The marginal case $\alpha=3$ is obtained by analytic continuation $\alpha=3-\epsilon$ and yields the logarithm \eqref{eq:LogKernelIdentity}:
\begin{equation}
\int \frac{d^3\mathbf{k}}{(2\pi)^3}\,\frac{e^{i\mathbf{k}\cdot\mathbf{r}}}{k^{3}}
= \lim_{\epsilon\to 0^+} \frac{\Gamma\!\big(\tfrac{\epsilon}{2}\big)}{2^{3-\epsilon}\,\pi^{3/2}\,\Gamma\!\big(\tfrac{3-\epsilon}{2}\big)}\,r^{-\epsilon}
= -\,\frac{1}{2\pi^2}\,\ln(\mu r)\,,
\end{equation}
where the scale $\mu$ collects scheme-dependent constants into the argument of the logarithm. Finally, combining \eqref{eq:1overk2} and \eqref{eq:LogKernelIdentity} reproduces \eqref{eq:PhiMarginal}.

\section{From the covariant nonlocal action to the Newtonian operator equation}
\label{app:non-local_action}

We start from the covariant action quoted in the main text,
\begin{equation}
\begin{split}
&S \;=\; \frac{1}{16\pi G_N}\int d^4x\,\sqrt{-g}\,R\\&\qquad
\;+\;\frac{k_*}{16\pi G_N}\int d^4x\,\sqrt{-g}\; R\,P_s\,(-\Box)^{-1/2}R,
\label{eq:A-action}
\end{split}
\end{equation}
where $P_s$ projects onto the scalar (spin-0) sector of metric fluctuations and $(-\Box)^{-1/2}$
is a nonlocal, self-adjoint operator. We linearize around Minkowski space
$g_{\mu\nu}=\eta_{\mu\nu}+h_{\mu\nu}$, adopt harmonic gauge $\partial^\mu \bar h_{\mu\nu}=0$ with
$\bar h_{\mu\nu}:=h_{\mu\nu}-\tfrac12 \eta_{\mu\nu} h$, and focus on the weak, static limit relevant
for the Newtonian potential $\Phi$ via $h_{00}=-2\Phi$. For static sources ($\partial_0=0$,
$T_{00}\simeq \rho$), the d'Alembertian reduces to the Laplacian, $-\Box \to -\nabla^2$.\vspace{2mm}

Varying \eqref{eq:A-action} yields field equations of the form
\begin{equation}
G_{\mu\nu} \;+\; k_*\,\mathcal H_{\mu\nu} \;=\; 8\pi G_N\,T_{\mu\nu},
\label{eq:A-feq}
\end{equation}
where $\mathcal H_{\mu\nu}$ denotes the contribution from the nonlocal term. At linear order about
flat space, the modification affects only the scalar (trace) sector because of the projector $P_s$.
Taking the $00$-component and going to Fourier space for static fields ($k^0=0$), one finds
\begin{equation}
-\,k^2\,\Phi(\mathbf k) \;=\; 4\pi\,G_N\!\left[\,1+\frac{k_*}{k}\,\right]\rho(\mathbf k),
\label{eq:A-kspace}
\end{equation}
where $k=|\mathbf k|$. This is the modified Poisson relation in momentum space. Equivalently, in
position space it corresponds to the operator-valued Poisson equation
\begin{equation}
\nabla^2 \Phi(\mathbf r) \;=\; 4\pi\, G(-\nabla^2)\,\rho(\mathbf r),
\end{equation}
where
\begin{equation}
G(-\nabla^2) \;\equiv\; G_N\!\left[\,1 + k_*\,(-\nabla^2)^{-1/2}\right].
\label{eq:A-operator-poisson}
\end{equation}
Equations \eqref{eq:A-kspace}–\eqref{eq:A-operator-poisson} are precisely Eqs.~(28)–(29) in the
main text.\vspace{2mm}

For completeness, the fractional operator can be defined in terms of the d'Alembertian as
\begin{equation}
(-\Box)^{-1/2}
\;=\;
\frac{1}{\Gamma(\tfrac12)}\int_0^\infty \! ds\, s^{-1/2}\, e^{\,s\Box}
\;=\;
\int_0^\infty \!\frac{d\mu^2}{\pi\,\mu}\,\frac{1}{-\Box+\mu^2},
\label{eq:A-frac-def}
\end{equation}
where the second representation exhibits $(-\Box)^{-1/2}$ as a superposition of massive,
Yukawa-type resolvents with positive spectral weight. In the static limit, $-\Box\to -\nabla^2$ and
\eqref{eq:A-frac-def} reduces to the Fourier multiplier $k^{-1}$ used above. With the retarded
prescription for each resolvent $(-\Box+\mu^2)^{-1}$, the construction ensures causal response.\vspace{2mm}

Thus the nonlocal action \eqref{eq:A-action} reproduces the momentum-space running
$G(k)=G_N[1+k_*/k]$ in the static Newtonian limit, and the corresponding position-space operator
equation \eqref{eq:A-operator-poisson} follows directly.

\bibliography{IR_bib}

@article{Reuter:2004nv,
    author = "Reuter, M. and Weyer, H.",
    title = "{Running Newton constant, improved gravitational actions, and galaxy rotation curves}",
    eprint = "hep-th/0410117",
    archivePrefix = "arXiv",
    reportNumber = "MZ-TH-04-14",
    doi = "10.1103/PhysRevD.70.124028",
    journal = "Phys. Rev. D",
    volume = "70",
    pages = "124028",
    year = "2004"
}

@article{Cirelli:2024ssz,
    author = "Cirelli, Marco and Strumia, Alessandro and Zupan, Jure",
    title = "{Dark Matter}",
    eprint = "2406.01705",
    archivePrefix = "arXiv",
    primaryClass = "hep-ph",
    month = "6",
    year = "2024",
    journal=""
}

@article{Arbey:2021gdg,
    author = "Arbey, A. and Mahmoudi, F.",
    title = "{Dark matter and the early Universe: a review}",
    eprint = "2104.11488",
    archivePrefix = "arXiv",
    primaryClass = "hep-ph",
    reportNumber = "CERN-TH-2021-066",
    doi = "10.1016/j.ppnp.2021.103865",
    journal = "Prog. Part. Nucl. Phys.",
    volume = "119",
    pages = "103865",
    year = "2021"
}

@article{Grumiller:2010bz,
    author = "Grumiller, Daniel",
    title = "{Model for gravity at large distances}",
    eprint = "1011.3625",
    archivePrefix = "arXiv",
    primaryClass = "astro-ph.CO",
    reportNumber = "TUW-10-19",
    doi = "10.1103/PhysRevLett.106.039901",
    journal = "Phys. Rev. Lett.",
    volume = "105",
    pages = "211303",
    year = "2010",
    note = "[Erratum: Phys.Rev.Lett. 106, 039901 (2011)]"
}

@article{Drlica-Wagner:2022lbd,
    author = "Drlica-Wagner, Alex and others",
    title = "{Report of the Topical Group on Cosmic Probes of Dark Matter for Snowmass 2021}",
    eprint = "2209.08215",
    archivePrefix = "arXiv",
    primaryClass = "hep-ph",
    reportNumber = "FERMILAB-FN-1211-PPD",
    doi = "10.2172/1898829",
    month = "9",
    year = "2022",
    journal=""
}

@article{milgrom1983modification,
  title={A modification of the Newtonian dynamics as a possible alternative to the hidden mass hypothesis},
  author={Milgrom, Mordehai},
  journal={Astrophysical Journal, Part 1 (ISSN 0004-637X), vol. 270, July 15, 1983, p. 365-370. Research supported by the US-Israel Binational Science Foundation.},
  volume={270},
  pages={365--370},
  year={1983}
}

@article{milgrom1983modification1,
  title={A modification of the Newtonian dynamics-Implications for galaxies},
  author={Milgrom, Mordehai},
  journal={Astrophysical Journal, Part 1 (ISSN 0004-637X), vol. 270, July 15, 1983, p. 371-383.},
  volume={270},
  pages={371--383},
  year={1983}
}

@article{Bertone:2004pz,
    author = "Bertone, Gianfranco and Hooper, Dan and Silk, Joseph",
    title = "{Particle dark matter: Evidence, candidates and constraints}",
    eprint = "hep-ph/0404175",
    archivePrefix = "arXiv",
    reportNumber = "FERMILAB-PUB-04-047-A",
    doi = "10.1016/j.physrep.2004.08.031",
    journal = "Phys. Rept.",
    volume = "405",
    pages = "279--390",
    year = "2005"
}

@article{Perivolaropoulos:2019vgl,
    author = "Perivolaropoulos, L. and Skara, F.",
    title = "{Reconstructing a Model for Gravity at Large Distances from Dark Matter Density Profiles}",
    eprint = "1903.06554",
    archivePrefix = "arXiv",
    primaryClass = "gr-qc",
    doi = "10.1103/PhysRevD.99.124006",
    journal = "Phys. Rev. D",
    volume = "99",
    number = "12",
    pages = "124006",
    year = "2019"
}

@article{Das:2022ozi,
    author = "Das, Saurya and Sur, Sourav",
    title = "{Dark matter or strong gravity?}",
    eprint = "2205.07153",
    archivePrefix = "arXiv",
    primaryClass = "gr-qc",
    doi = "10.1142/S0218271822420202",
    journal = "Int. J. Mod. Phys. D",
    volume = "31",
    number = "14",
    pages = "2242020",
    year = "2022"
}

@article{Das:2023gpf,
    author = "Das, Saurya and Sur, Sourav",
    title = "{Gravitational lensing and missing mass}",
    eprint = "2303.03259",
    archivePrefix = "arXiv",
    primaryClass = "gr-qc",
    doi = "10.1016/j.physo.2023.100150",
    journal = "Phys. Open",
    volume = "15",
    pages = "100150",
    year = "2023"
}

@article{Mistele:2024hfh,
    author = "Mistele, Tobias and McGaugh, Stacy and Lelli, Federico and Schombert, James and Li, Pengfei",
    title = "{Indefinitely Flat Circular Velocities and the Baryonic Tully{\textendash}Fisher Relation from Weak Lensing}",
    eprint = "2406.09685",
    archivePrefix = "arXiv",
    primaryClass = "astro-ph.GA",
    doi = "10.3847/2041-8213/ad54b0",
    journal = "Astrophys. J. Lett.",
    volume = "969",
    number = "1",
    pages = "L3",
    year = "2024"
}

@inproceedings{tohline1983stabilizing,
  title={Stabilizing a cold disk with a 1/r force law},
  author={Tohline, Joel E},
  booktitle={Symposium-International Astronomical Union},
  volume={100},
  pages={205--206},
  year={1983},
  organization={Cambridge University Press}
}

@article{kuhn1987non,
  title={Non-Newtonian forces and the invisible mass problem},
  author={Kuhn, JR and Kruglyak, L},
  journal={Astrophysical Journal, Part 1 (ISSN 0004-637X), vol. 313, Feb. 1, 1987, p. 1-12. NSF-supported research.},
  volume={313},
  pages={1--12},
  year={1987}
}

@inproceedings{bekenstein1988missing,
  title={The missing light puzzle: a hint about gravitation?},
  author={Bekenstein, JD},
  booktitle={Proceedings of the 2nd Canadian Conference on General Relativity and Relativistic Astrophysics},
  pages={68--104},
  year={1988}
}

@inproceedings{Mashhoon:2011mb,
    author = "Mashhoon, Bahram",
    title = "{Nonlocal Gravity}",
    booktitle = "{14th Brazilian School of Cosmology and Gravitation}",
    eprint = "1101.3752",
    archivePrefix = "arXiv",
    primaryClass = "gr-qc",
    month = "1",
    year = "2011"
}

@article{Donoghue:1994dn,
    author = "Donoghue, John F.",
    title = "{General relativity as an effective field theory: The leading quantum corrections}",
    eprint = "gr-qc/9405057",
    archivePrefix = "arXiv",
    reportNumber = "UMHEP-408",
    doi = "10.1103/PhysRevD.50.3874",
    journal = "Phys. Rev. D",
    volume = "50",
    pages = "3874--3888",
    year = "1994"
}

@article{LopezNacir:2006tn,
    author = "Lopez Nacir, D. and Mazzitelli, F. D.",
    title = "{Running of Newton's constant and non integer powers of the d'Alembertian}",
    eprint = "hep-th/0610031",
    archivePrefix = "arXiv",
    doi = "10.1103/PhysRevD.75.024003",
    journal = "Phys. Rev. D",
    volume = "75",
    pages = "024003",
    year = "2007"
}

@article{Sofue:2015tsa,
    author = "Sofue, Yoshiaki",
    title = "{Rotation curve decomposition for size{\textendash}mass relations of bulge, disk, and dark halo components in spiral galaxies}",
    eprint = "1510.05752",
    archivePrefix = "arXiv",
    primaryClass = "astro-ph.GA",
    doi = "10.1093/pasj/psv103",
    journal = "Publ. Astron. Soc. Jap.",
    volume = "68",
    number = "1",
    pages = "2",
    year = "2016"
}

@article{Farooq:2016zwm,
    author = "Farooq, Omer and Madiyar, Foram Ranjeet and Crandall, Sara and Ratra, Bharat",
    title = "{Hubble Parameter Measurement Constraints on the Redshift of the Deceleration{\textendash}acceleration Transition, Dynamical Dark Energy, and Space Curvature}",
    eprint = "1607.03537",
    archivePrefix = "arXiv",
    primaryClass = "astro-ph.CO",
    doi = "10.3847/1538-4357/835/1/26",
    journal = "Astrophys. J.",
    volume = "835",
    number = "1",
    pages = "26",
    year = "2017"
}

@article{Foreman-Mackey:2012any,
    author = "Foreman-Mackey, Daniel and Hogg, David W. and Lang, Dustin and Goodman, Jonathan",
    title = "{emcee: The MCMC Hammer}",
    eprint = "1202.3665",
    archivePrefix = "arXiv",
    primaryClass = "astro-ph.IM",
    doi = "10.1086/670067",
    journal = "Publ. Astron. Soc. Pac.",
    volume = "125",
    pages = "306--312",
    year = "2013"
}

@article{Planck:2018vyg,
    author = "Aghanim, N. and others",
    collaboration = "Planck",
    title = "{Planck 2018 results. VI. Cosmological parameters}",
    eprint = "1807.06209",
    archivePrefix = "arXiv",
    primaryClass = "astro-ph.CO",
    doi = "10.1051/0004-6361/201833910",
    journal = "Astron. Astrophys.",
    volume = "641",
    pages = "A6",
    year = "2020",
    note = "[Erratum: Astron.Astrophys. 652, C4 (2021)]"
}

@article{Copi:2003xd,
    author = "Copi, Craig J. and Davis, Adam N. and Krauss, Lawrence M.",
    title = "{A New nucleosynthesis constraint on the variation of G}",
    eprint = "astro-ph/0311334",
    archivePrefix = "arXiv",
    reportNumber = "CWRU-P35-03",
    doi = "10.1103/PhysRevLett.92.171301",
    journal = "Phys. Rev. Lett.",
    volume = "92",
    pages = "171301",
    year = "2004"
}
\bibliographystyle{utphys1}
\end{document}